\documentclass[aps,prl,twocolumn,superscriptaddress,showpacs,letter]{revtex4-1}

\usepackage{graphicx}
\usepackage{dcolumn}
\usepackage{bm}
\usepackage{cancel}
\usepackage{color}

\newcommand{\bnfa}{Ba$_{1-x}$Na$_x$Fe$_2$As$_2$}
\newcommand{\bkfa}{Ba$_{1-x}$K$_x$Fe$_2$As$_2$}
\newcommand{\bfa}{BaFe$_2$As$_2$}

\newcommand{\ef}{$E_F$}
\newcommand{\kf}{$k_F$}

\newcommand{\kone}{$k_1$}
\newcommand{\ktwo}{$k_2$}
\newcommand{\qone}{$\mathbf{q_1}$}
\newcommand{\qtwo}{$\mathbf{q_2}$}
\newcommand{\qt}{$\mathbf{q_3}$}
\newcommand{\tc}{$T_c$}
\newcommand{\ts}{$T_S$}
\newcommand{\tdq}{$T_{DQ}$}
\newcommand{\tn}{$T_N$}
\newcommand{\cfour}{C$_4$}
\newcommand{\ctwo}{C$_2$}
\newcommand{\dxy}{$d_{xy}$}
\newcommand{\dxz}{$d_{xz}$}
\newcommand{\dyz}{$d_{yz}$}

\newcommand{\g}{$\Gamma$}
\newcommand{\gx}{$\Gamma-X$}
\newcommand{\gy}{$\Gamma-Y$}

\begin{document}

\title{Spectral Evidence for Emergent Order in \bnfa}

\author{M. Yi}
\email{mingyi@berkeley.edu; mingyi@rice.edu}
\affiliation{Department of Physics, University of California Berkeley, Berkeley, CA 94720, USA}
\affiliation{Department of Physics and Astronomy, Rice University, Houston, TX 77005, USA}
\author{A. Frano}
\affiliation{Department of Physics, University of California Berkeley, Berkeley, CA 94720, USA}
\affiliation{Advanced Light Source, Lawrence Berkeley National Lab, Berkeley, CA 94720, USA}
\author{D. H. Lu}
\affiliation{Stanford Synchrotron Radiation Lightsource, SLAC National Accelerator Laboratory, Menlo Park, CA 94025, USA}
\author{Y. He}
\affiliation{Stanford Institute of Materials and Energy Sciences, Stanford University, Stanford, CA 94305, USA}
\affiliation{Departments of Physics and Applied Physics, and Geballe Laboratory for Advanced Materials, Stanford University, Stanford, CA 94305, USA}
\author{Meng Wang}
\affiliation{School of Physics, Sun Yat-Sen University, Guangzhou 510275, China}
\author{B. A. Frandsen}
\affiliation{Materials Sciences Division, Lawrence Berkeley National Laboratory, Berkeley, CA 94720, USA}
\author{A. F. Kemper}
\affiliation{Department of Physics, North Carolina State University, Raleigh, NC 27695, USA}
\author{R. Yu}
\affiliation{Department of Physics, Renmin University of China, Beijing 100872, China}
\author{Q. Si}
\affiliation{Department of Physics and Astronomy, Rice University, Houston, TX 77005, USA}
\author{L. Wang}
\affiliation{Institute for Solid State Physics, Karlsruhe Institute of Technology, 76021 Karlsruhe, Germany}
\affiliation{Kirchhoff-Institute for Physics, Universität Heidelberg, D-69120 Heidelberg, Germany}
\author{M. He}
\author{F. Hardy}
\author{P. Schweiss}
\author{P. Adelmann}
\author{T. Wolf}
\affiliation{Institute for Solid State Physics, Karlsruhe Institute of Technology, 76021 Karlsruhe, Germany}
\author{M. Hashimoto}
\affiliation{Stanford Synchrotron Radiation Lightsource, SLAC National Accelerator Laboratory, Menlo Park, CA 94025, USA}
\author{S.-K. Mo}
\author{Z. Hussain}
\affiliation{Advanced Light Source, Lawrence Berkeley National Lab, Berkeley, CA 94720, USA}
\author{M. Le Tacon}
\affiliation{Institute for Solid State Physics, Karlsruhe Institute of Technology, 76021 Karlsruhe, Germany}
\author{A. E. B\"ohmer}
\affiliation{Institute for Solid State Physics, Karlsruhe Institute of Technology, 76021 Karlsruhe, Germany}
\author{D.-H. Lee}
\affiliation{Department of Physics, University of California Berkeley, Berkeley, CA 94720, USA}
\affiliation{Materials Sciences Division, Lawrence Berkeley National Laboratory, Berkeley, CA 94720, USA}
\author{Z.-X. Shen}
\affiliation{Stanford Institute of Materials and Energy Sciences, Stanford University, Stanford, CA 94305, USA}
\affiliation{Departments of Physics and Applied Physics, and Geballe Laboratory for Advanced Materials, Stanford University, Stanford, CA 94305, USA}
\author{C. Meingast}
\affiliation{Institute for Solid State Physics, Karlsruhe Institute of Technology, 76021 Karlsruhe, Germany}
\author{R. J. Birgeneau}
\email{robertjb@berkeley.edu}
\affiliation{Department of Physics, University of California Berkeley, Berkeley, CA 94720, USA}
\affiliation{Materials Sciences Division, Lawrence Berkeley National Laboratory, Berkeley, CA 94720, USA}
\affiliation{Department of Materials Science and Engineering, University of California, Berkeley, CA 94720, USA}

\date{\today}

\begin{abstract}
We report an angle-resolved photoemission spectroscopy study of the iron-based superconductor family, \bnfa. This system harbors the recently discovered double-Q magnetic order appearing in a reentrant \cfour~phase deep within the underdoped regime of the phase diagram that is otherwise dominated by the coupled nematic phase and collinear antiferromagnetic order. From a detailed temperature-dependence study, we identify the electronic response to the nematic phase in an orbital-dependent band shift that strictly follows the rotational symmetry of the lattice and disappears when the system restores \cfour~symmetry in the low temperature phase. In addition, we report the observation of a distinct electronic reconstruction that cannot be explained by the known electronic orders in the system. 
\end{abstract}


\maketitle


\begin{figure}
\includegraphics[width=0.5\textwidth]{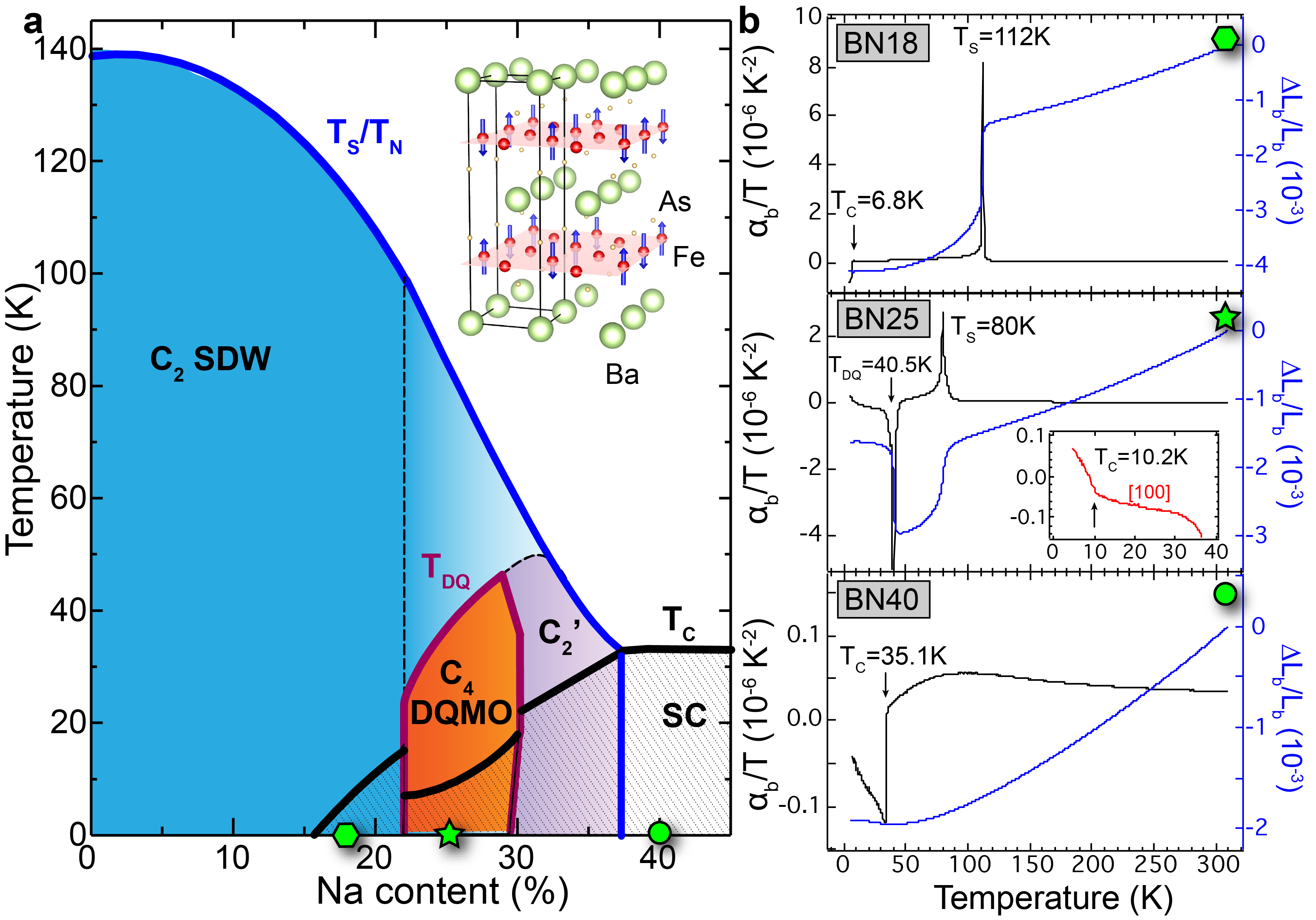}
\caption{\label{fig:fig1} \bnfa~phase diagram and dilatometry measurements. (a) Phase diagram adapted from Ref. [9]. The inset shows the DQMO in the reentrant \cfour~phase. (b) Temperature-dependence of the thermal expansion coefficient, $\alpha$, plotted as $\alpha/T$, and relative thermal expansion, $\Delta L_b/L_b$, for the samples measured by ARPES. The inset shows $\alpha/T$ measured along the twinned ([100]) direction of BN25.}
\end{figure}

Quantum phases emerge in strongly correlated electron systems via the interplay of four fundamental degrees of freedom: lattice, charge, orbital, and spin. For both copper- and iron-based high temperature superconductors, the spin degree of freedom plays a prominent role as long-range magnetism has been found in the parent phase of both materials~\cite{Keimer2015,Dai2015}. In the iron-based superconductors (FeSCs), magnetic order has largely been discussed in relation to the orbital degrees of freedom after the discovery of the electronic nematic phase~\cite{Chu2010,Tanatar2010} and the associated orbital symmetry breaking~\cite{Yi2011}, which is accompanied by the \cfour~rotational symmetry breaking through the tetragonal to orthorhombic structural transition. The coupled collinear antiferromagnetic order (CAF) and the nematic order persist throughout the underdoped region of many FeSC families. Very recently, however, a reentrant \cfour~magnetic phase has been discovered in the underdoped region of many hole-doped FeSCs ((Ba,Sr,Ca)Fe$_2$As$_2$ doped with Na/K) close to the optimal doping where \tc~is maximal~\cite{Avci2014,Allred2016,Waber2015,Wang2016,Bohmer2015,Taddei2017}. Within this reentrant \cfour~ phase, tetragonal symmetry is restored while the spin order persists and reconstructs, suggesting the interesting possibility of the decoupling of spin order and nematic order. M\"ossbauer~\cite{Allred2016} and neutron diffraction~\cite{Waber2015} measurements together demonstrate that the magnetic order in the reentrant \cfour~ phase is of a double-Q type, where the moments point along the c-axis and follow the superposition of two spin density waves along \qone~= $(\pi/2, \pi/2, \pi)_T$ and \qtwo~= $(\pi/2, -\pi/2, \pi)_T$ in the tetragonal 2-Fe Brillouin zone notation, and became known as the double-Q magnetic order (DQMO). Spatially, this spin structure can be viewed as two Fe sublattices where one sublattice is antiferromagnetic while the other is non-magnetic, respecting \cfour~rotational symmetry (inset of Fig.~\ref{fig:fig1}a). 

To elucidate the interactions between the rich electronic orders in this system, we study the reentrant \cfour~phase in the \bnfa~family using angle-resolved photoemission spectroscopy (ARPES). Three doping regimes in the \bnfa~family are studied, x=0.18 (BN18), 0.25 (BN25), and 0.4 (BN40), representing the \ctwo~phase-only regime, the reentrant \cfour~phase regime, and the purely superconducting phase regime, respectively (Fig.~\ref{fig:fig1}). Prominent orbital anisotropy is observed in the nematic phase of BN18, consistent with the understanding of nematic order in other FeSCs. For BN25, we observe a similar anisotropic orbital-dependent band shift that onsets as the system enters the nematic phase marked by \ts, and disappears when the system enters the reentrant \cfour~phase at lower temperatures. In addition, we observe a distinct electronic reconstruction exhibiting a different temperature evolution. This new electronic reconstruction cannot be explained by any of the known electronic orders in the system, including nematicity, CAF, and DQMO. The absence of this type of electronic reconstruction in the other doping regimes strongly suggests that this order arises from a coupling to the DQMO, potentially revealing a parallelism akin to the coupling between the nematic phase and CAF prevalent in the underdoped regime of FeSCs.

\begin{figure}
\includegraphics[width=0.5\textwidth]{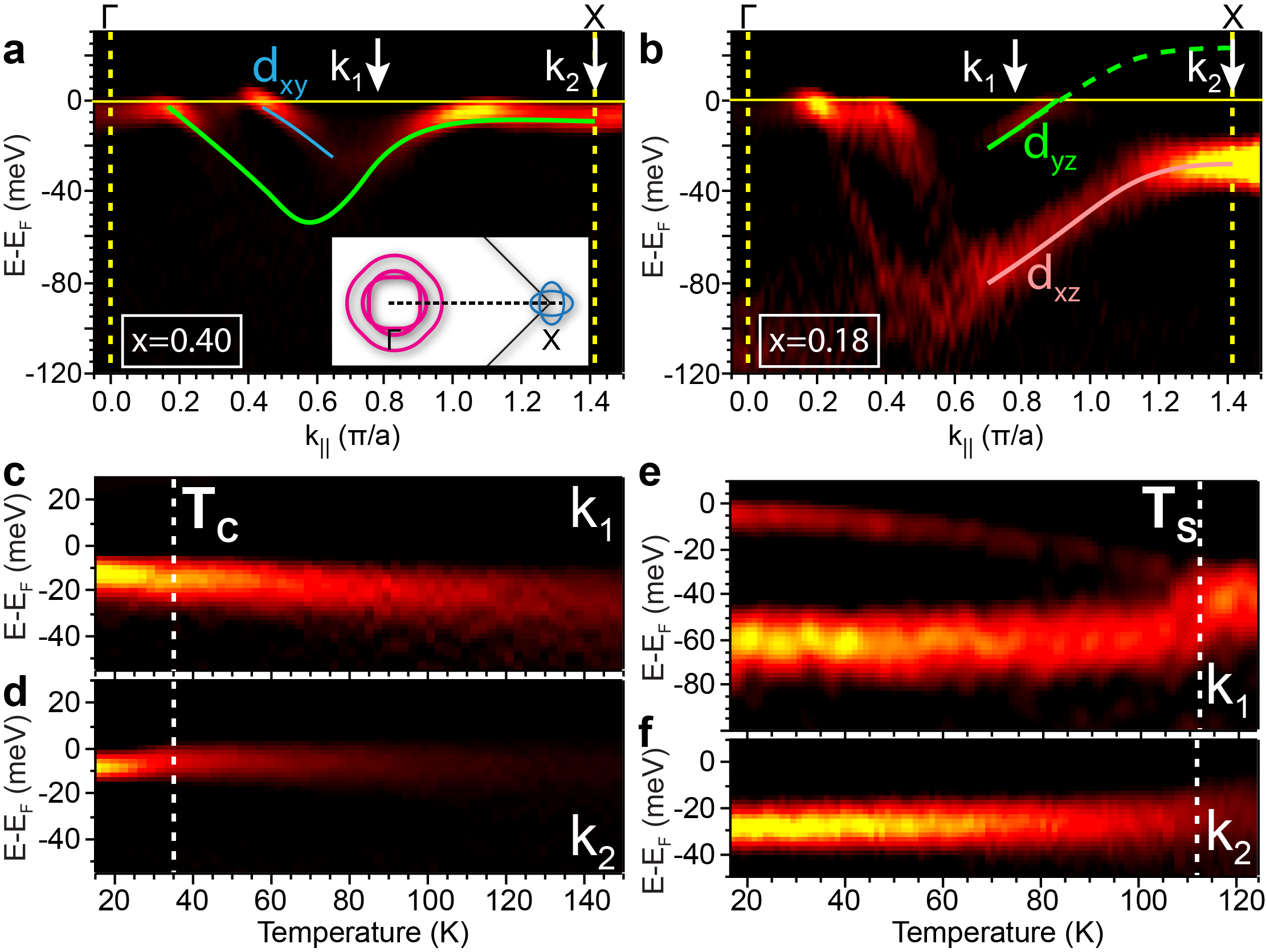}
\caption{\label{fig:fig4}Temperature dependence of BN40 and BN18. Second derivative of the measured band dispersions along the \gx~direction of BN40, taken at 15K. (b) Same measurement for the twinned orthorhombic phase of BN18 at 15K. (c)-(d) Fine temperature dependence at the selected momenta on BN40. (e)-(f) Fine temperature dependence at the selected momenta on BN18. All measurements were taken with 25eV photons under polarization odd with respect to the cut direction.}
\end{figure}

\begin{figure}
\includegraphics[width=0.5\textwidth]{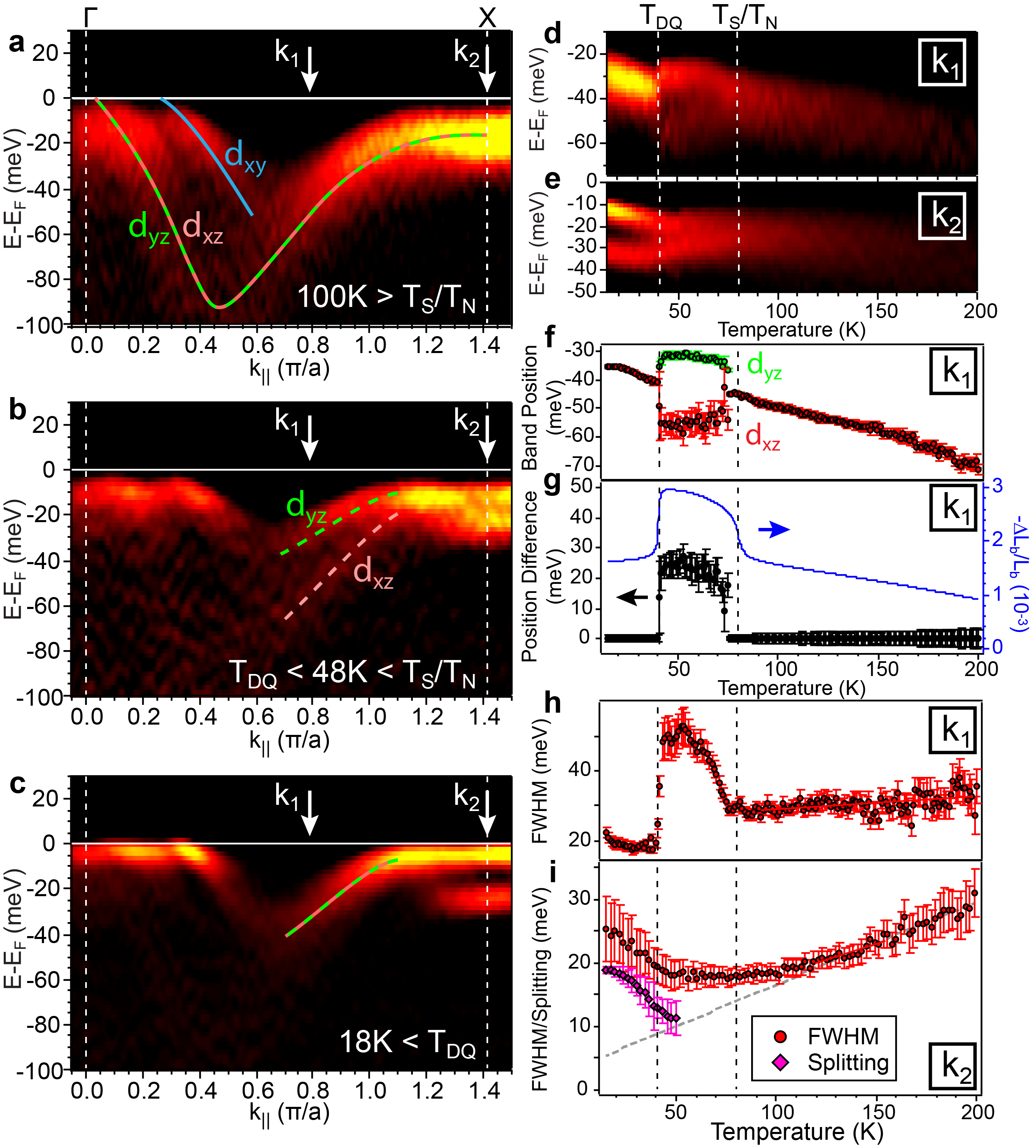}
\caption{\label{fig:fig2}Temperature dependence of BN25. Band dispersions are shown along $\Gamma-X$ for (a) T $>$ \ts~= \tn, (b) \ts~= \tn~$>$ T $>$ \tdq, and (c) T $<$ \tdq. Temperature-dependent EDCs are shown for \kone~(d) and \ktwo~(e) as marked by arrows in (a). All image plots are second energy derivatives. (f) Fitted band positions of the temperature dependence at \kone~using one peak for the \cfour~phases (T $<$ \tdq~and T $>$ \ts) and two peaks for the \ctwo~phase (\tdq $<$ T $<$ \ts). (g) Band splitting (black) is compared to the lattice distortion measured by dilatometry (blue). Each second derivative EDC in (d)-(e) is fitted by a constant background and a single Gaussian peak where anomalous broadening suggests band splitting~\cite{C4ARPES_SM}. Fitted single-peak FWHM for \kone~(h) and \ktwo~(i) shown as a function of temperature. The black line for \ktwo~is a fitted linear background extrapolating the behavior of a system without any ordering~\cite{C4ARPES_SM}. The low temperature EDCs in (e) are also fitted by two peaks, where the extracted splitting size is shown in magenta in (i)~\cite{C4ARPES_SM}. All measurements were taken with 25eV photons under LH polarization.}
\end{figure}

High quality single crystals of \bnfa~were grown using the self-flux method~\cite{Wang2016}, and individually characterized by thermal expansion measurements via dilatometry~\cite{Bohmer2016} (Fig.~\ref{fig:fig1}b). Throughout this Letter, we refer to the transition temperatures of the structural distortion, CAF order, DQMO, and superconductivity as \ts, \tn, \tdq, and \tc, respectively. BN18 in the very underdoped regime undergoes the structural and magnetic transitions at \ts~= \tn~= 112K, and superconducts at \tc~= 6.8K. BN25 first goes through the structural/magnetic transitions at \ts~= \tn~= 80K, then enters the reentrant \cfour~magnetic phase at \tdq~= 40.5K, and finally superconducts at \tc~= 10.2K. BN40 only exhibits superconductivity onsetting at \tc~= 35.1K. ARPES measurements were carried out at beamlines 5-4 and 5-2 of the Stanford Synchrotron Radiation Lightsource and beamline 10.0.1 of the Advanced Light Source using SCIENTA R4000 electron analyzers. The total energy resolution was set to 10 meV or better and the angular resolution was 0.3$^o$. Single crystals were cleaved in-situ at 10 K. All measurements were done in ultrahigh vacuum with a base pressure lower than 4x10-11 torr. The simulations were produced based on a 3-dimensional five-orbital tight-binding model of \bfa~from Density Functional Theory (DFT) band structure~\cite{Graser2010}. To match roughly the observed band structure of BN25, the DFT band structure for undoped \bfa~was shifted up by 0.12eV in energy to account for the hole-doping, and then renormalized by a factor of 4.3. For consistency, we use the tetragonal 2-Fe notation, where the lattice constants for BN25 at 300K are $a_T$ = $b_T$ = 3.921\AA, and $c_T$ = 13.110\AA. 

We begin with the simplest compound, BN40, which has no symmetry breaking phases except superconductivity (Fig.~\ref{fig:fig4}a). The measured band dispersions are very similar to those of the widely studied optimally hole-doped FeSC, \bkfa~\cite{Lin2008,Zhang2010,Xu2011}. Under $s$ polarization, two hole bands are visible near \g. The one with a larger momentum crossing, \kf, is of \dxy~orbital character~\cite{Yi2011}. The one with a smaller \kf~disperses and upturns into an intense, flat, hole-like feature towards the X point. This band is predominantly of \dyz~character along \gx, and by \cfour~symmetry, \dxz~along \gy. No anomalous band reconstruction is observed in the temperature dependence (Fig.~\ref{fig:fig4}c-d). Next we examine BN18. Since this sample is unstressed, the development of structural domains in the nematic phase allows us to observe both orthogonal directions simultaneously. In the normal state, the \dyz~band is degenerate with its counterpart, the \dxz~band in the orthogonal direction. When \cfour~symmetry is broken at the onset of the nematic phase, this degeneracy is lifted with an upward (downward) shift of the \dyz~(\dxz) band in orthogonal directions. On a twinned sample, this orbital-dependent band shift is therefore manifested in a band splitting (Fig.~\ref{fig:fig4}b)---a hallmark of the orbital anisotropy associated with the electronic nematic order~\cite{Yi2011}. This splitting onsets clearly at \ts~(Fig.~\ref{fig:fig4}e). We note that in order to see this, we must go to a momentum in between \g~and X, such as \kone, as the \dyz~band shifts to above \ef~at the X point, making it unobservable (Fig.~\ref{fig:fig4}f).

Next, we discuss the sample harboring the reentrant \cfour~ phase, BN25. In Fig.~\ref{fig:fig2} we show the bands along the \gx~direction in the three distinct temperature regimes, i) the tetragonal and paramagnetic normal state ($T >$ \ts~= \tn), ii) the orthorhombic CAF phase (\ts~= \tn~$> T >$ \tdq), and iii) the tetragonal double-Q magnetic phase (T $<$ \tdq). The normal state dispersions are similar to those of BN40 (Fig.~\ref{fig:fig2}a). When cooled below \ts, a splitting of the \dyz/\dxz~band appears analogously to the BN18 sample indicative of the orbital anisotropy that appears in the nematic phase (Fig.~\ref{fig:fig2}b). When the reentrant \cfour~phase is entered (Fig.~\ref{fig:fig2}c), this splitting disappears, reflecting the restored \cfour~rotational symmetry. From a fine temperature-dependent measurement at a momentum between \g~and X as we have done for BN18, \kone, (Fig.~\ref{fig:fig2}d), we see clearly a single feature splitting into two near \ts~and sharply merging back into one at \tdq~in a strongly first-order manner, reminiscent of the behavior of structural Bragg peaks measured by powder neutron diffraction~\cite{Avci2014}. As a visualization of the orbital anisotropy order parameter, we fit the band positions as a function of temperature, fitting one peak for the \cfour~phases and two peaks for the \ctwo~phase (Fig.~\ref{fig:fig2}f). The extracted behavior of the splitting size agrees well with the behavior of the lattice distortion measured by dilatometry (Fig.~\ref{fig:fig2}g). While the number of peaks used in this fitting method is based on the number of bands expected in accordance with the crystal symmetry in different temperature regimes, another unbiased way to determine where the phase transitions occur is to plot the full width at half maximum (FWHM) of a single peak fit for all temperatures~\cite{C4ARPES_SM}. Indeed, an anomalous broadening of this fitted FWHM (Fig.~\ref{fig:fig2}h) is observed precisely between \ts~and \tdq, in good agreement with both the lattice distortion and orbital anisotropy extracted from the band splitting.

\begin{figure}
\includegraphics[width=0.5\textwidth]{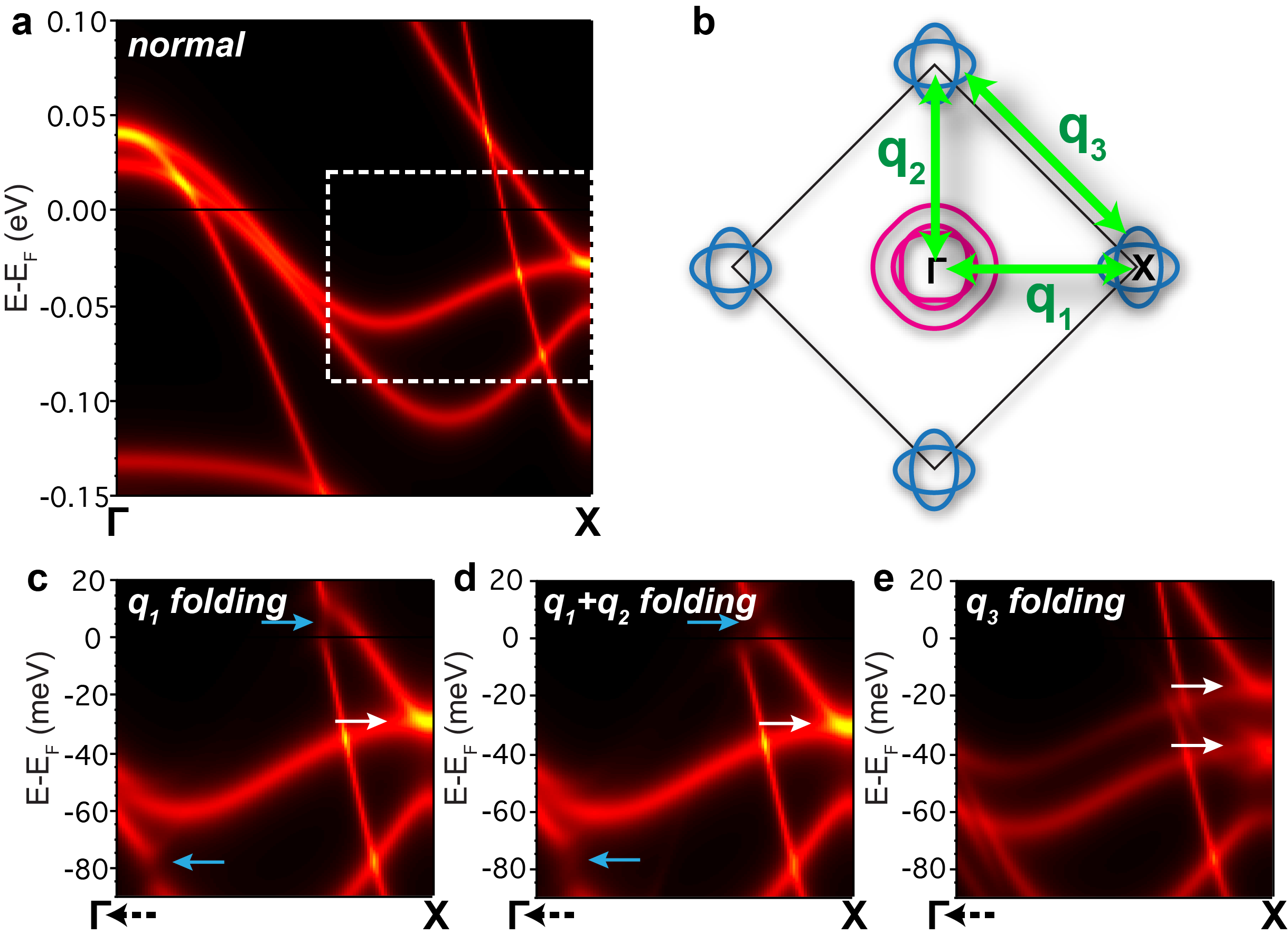}
\caption{\label{fig:fig3}Simulated effects for different electronic orders. A tight-binding fit of undoped \bfa~to DFT band structure is used, with an overall shift of 0.12eV and renormalization of 4.3 to best match the measured dispersions in the normal state of BN25. (a) Calculated dispersions along $\Gamma-X$ in the normal state. (b) Schematic of the FS showing the relevant folding vectors. Simulated reconstructed bands in the box region in (a) are shown for (c) the CAF order with \qone~= $(\pi/2, \pi/2, \pi)_T$, (d) the double-Q order with both \qone~= $(\pi/2, \pi/2, \pi)_T$ and \qtwo~= $(\pi/2, -\pi/2, \pi)_T$, and (e) the CO with \qt~= $(\pi, 0, 0)_T$, where the doubling of bands is observed around X (white arrows). Blue arrows point to magnetic energy gaps.}
\end{figure}

Having demonstrated that the electronic structure of BN25 follows the onset and disappearance of the nematic order through the reentrant \cfour~phase, we now point to an unexpected observation. In contrast to the non-monotonic temperature-dependence of the band splitting magnitude at \kone, a simultaneous measurement at the X point, \ktwo, reveals a strikingly distinct behavior---a single feature monotonically splitting into two with decreasing temperature (Fig.~\ref{fig:fig2}e). At first sight, one might also ascribe this splitting to orbital anisotropy onsetting at \ts. However, two considerations disprove this conclusion. First, the splitting at \ktwo~is largest in the reentrant \cfour~phase, where the electronic structure respects restored \cfour~rotational symmetry. Second, the evolution through \tdq~(Fig.~\ref{fig:fig2}e) is continuous, in contrast to the strongly first order transition at \kone~(Fig.~\ref{fig:fig2}d), suggesting a different mechanism than the one responsible for the splitting at ~\kone~. Hence, the splitting at ~\ktwo~cannot simply be due to the same orbital anisotropy that generates the splitting at ~\kone.

To extract the onset temperature of this splitting, we again exploit the single-peak fitting procedure used for \kone. The fitted FWHM as a function of temperature at \ktwo~ (Fig.~\ref{fig:fig2}i) shows an initial down slope that flattens and eventually upturns. From the control sample of BN40 where no orders exist above \tc, we see that the fitted FWHM at X narrows in a linear fashion with lowering temperature~\cite{C4ARPES_SM}. Hence, in comparison, while the broadening significantly takes off around \tdq~= 40.5K, the flattening at more elevated temperatures suggests a slow emergence of the splitting at X above \tdq, which is also visible in the temperature-dependence in Fig.~\ref{fig:fig2}e, as well as the clear band splitting at X in the 48K data taken above \tdq~(Fig.~\ref{fig:fig2}b). For the low temperature regime, the band splitting size can be reliably extracted from a two-peak fit (magenta in Fig.~\ref{fig:fig2}i), showing a rapid decrease upon raising the temperature approaching \tdq, and a non-zero presence at least 10K above this transition.

From the bands measured across the X point in the \cfour~phase (Fig.~\ref{fig:fig2}c), we see that the splitting at X results from the emergence of an additional band separated by 20meV from the original band. Since we have excluded its origin from being due to nematicity, we now explore whether it could be a result of band folding due to known magnetic orders in the system by performing a series of simulations incorporating the different magnetic orders into a three-dimensional tight binding model of the \bfa~normal state band structure calculated via DFT~\cite{Graser2010} (Fig.~\ref{fig:fig3}). We note that a more rigorous self-consistent calculation is needed to assess fully the effects of the intertwined orders in these materials, but we exploit this exercise to understand qualitatively the essential effects of the distinct electronic orders in these materials. We focus on the region near the X point (white box in Fig.~\ref{fig:fig3}a). First, for both the single-Q  \qone~= $(\pi/2, \pi/2, \pi)_T$ (Fig.~\ref{fig:fig3}c) and double-Q \qone~+ \qtwo~= $(\pi/2, \pi/2, \pi)_T$ + $(\pi/2, -\pi/2, \pi)_T$ (Fig.~\ref{fig:fig3}d) orders, the only kind of additional bands that can appear at X are bands that are folded from the \g~point~\cite{Yi2012,Zhang2012,Yi2014}. Since all the hole band tops at \g~are above \ef~and there are no bands at the energy where this new feature is observed, it cannot be a result of the folding via these known magnetic orders alone. Therefore, we have excluded all the known electronic orders in this system from being the possible origin of this electronic reconstruction. 

In contrast, to produce the band doubling effect at the X point as observed, one possibility is band folding associated with a checkerboard order with \qt~= $(\pi, 0, 0)_T$ (Fig.~\ref{fig:fig3}e), which indeed produces a band splitting at the X point. However, we note that a simple order of this q alone also cannot fully describe the data, where the intensity for the lower band of the doublet is only observable for a finite range of momenta around X, the cause of which cannot be captured in this simple simulation exercise. As this phase exists in the presence of the DQMO, a fully self-consistent calculation taking into consideration the combined effects of these intertwined orders may help bridge the precise comparison to the experimental observation. To understand the origin of this new order, we note that from neutron diffraction measured in the reentrant \cfour~phase, no new magnetic peak has been found at $\mathbf{q}$ = $(\pi, 0, 0)_T$~\cite{Waber2015}. Hence the order giving rise to this folding vector is unlikely to be magnetic in origin. We offer, instead, a different possible explanation of the splitting. Theoretical studies have suggested that a checkerboard charge order (CO) where the Fe sublattice with zero moment has different charge density than that of the antiferromagnetic sublattice would be compatible with the symmetry of the DQMO~\cite{Lorenzana2008,Fernandes2016,Gastiasoro2015}. Such a CO has a q-vector that is the sum of the two q-vectors of the DQMO, and could qualitatively and partially reproduce the type of band splitting we have observed. If this is indeed the CO, this result is reminiscent of the induced CO under a magnetic field observed in Fe$_x$Co$_{1-x}$TiO$_3$~\cite{Harris1997}. Observations by other probes in the reentrant \cfour~phase are also consistent with our results, including a phonon back-folding~\cite{Mallett2015}, as well as an electronic gap opening observed by Raman scattering with a similar characteristic energy scale as our observed band splitting~\cite{Yue2017}. 

Finally, a comparison between the three doping levels shows that the band splitting at X suggestive of the checkerboard order only occurs for BN25 (Fig.~\ref{fig:fig4}). Hence, the combined temperature and doping dependences suggest that the new order likely strongly couples to the DQMO. Theoretically, it is anticipated that an Ising-like checkerboard CO accompanies the DQMO ~\cite{Lorenzana2008,Fernandes2016,Gastiasoro2015} analogously as the Ising-nematic order to the single-Q CAF order, where the CO could either emerge simultaneously with the DQMO in a first order transition, or precede the formation of the DQMO~\cite{Fang2008,Xu2008,Dai2009,Chandra1990,Yu2017}. In the latter case, a Ginzburg-Landau analysis suggests that, when the DQMO develops upon lowering the temperature from a \ctwo~magnetic phase, the magnetic transition at \tdq~is first order, while the onset of the CO in the background of the Ising-nematic order could be continuous~\cite{C4ARPES_SM}. From our measurements, while we observe evidence of band splitting already emerging above \tdq~(Fig.~\ref{fig:fig2}b), the fast time-scale of the photoemission process does not allow us to preclude the possibility of this arising from fluctuation effects since the order parameter is seen to develop significantly below \tdq. However, our measurements do reveal strong order susceptibilities in this regime. Interestingly, strong nematic susceptibility is found inside the reentrant \cfour~phase by both measurements of the Young's modulus~\cite{Wang2018} as well as a pair distribution function measurement that observed local orthorhombicity deep in the reentrant \cfour~phase~\cite{Frandsen2017}, suggesting the strong first-order nature of the \tdq~transition as well as potential strong interaction of the nematic and charge fluctuations in this regime.

The finding of this emergent order and its strong coupling to the DQMO in underdoped FeSCs unveils another parallel with the intertwined electronic ``stripe" order  appearing at 1/8 doping in the La$_{2-x}$Ba$_x$CuO$_4$ cuprate system where the CO couples strongly to the spin order~\cite{Keimer2015,Tranquada1995}. Interestingly, in both the cuprate case~\cite{Keimer2015} and the \bnfa~case~\cite{Wang2016} reported here, the \tc~dome develops a suppression where the intertwined orders appear, suggesting a non-trivial interaction of these orders with superconductivity. Overall, our observation of the effects of the putative checkerboard electronic order in \bnfa~reveals that high temperature superconductivity in the FeSCs emerges in an elaborately intertwined regime where magnetism could potentially couple both to the orbital degrees of freedom and charge degrees of freedom, opening up exciting perspectives for theoretical investigations of the mechanism for high temperature superconductivity. Other measurements of the charge order, both direct and indirect, including especially by resonant x-ray scattering would help elucidate the nature of the couplings between the different orders and their fluctuations. We hope our report of the emergent order in the reentrant \cfour~phase could motivate further theoretical efforts in understanding the intricate electronic phase interactions in these rich material systems.

\begin{acknowledgments}
We thank Maria Gastiasoro, Keith Taddei, Yan Zhang, Yuan Li, and Fa Wang for fruitful discussions. Work at University of California, Berkeley and Lawrence Berkeley National Laboratory was funded by the U.S. Department of Energy, Office of Science, Office of Basic Energy Sciences, Materials Sciences and Engineering Division under Contract No. DE-AC02-05-CH11231 within the Quantum Materials Program (KC2202) and the Office of Basic Energy Sciences, U.S. DOE, Grant No. DE-AC03-76SF008. ARPES experiments were performed at the Stanford Synchrotron Radiation Lightsource and the Advanced Light Source, which are both operated by the Office of Basic Energy Sciences, U.S. DOE. Work at Renmin University is supported by the National Science Foundation of China Grant numbers 11374361 and 11674392 and Ministry of Science and Technology of China, National Program on Key Research Project Grant number 2016YFA0300504. Work at Rice University is supported by the U.S. Department of Energy, Office of Science, Basic Energy Sciences, under Award No. DE-SC0018197, and by the Robert A. Welch Foundation Grant No. C-1411. M.Y. acknowledges the L'Or\'eal For Women in Science Fellowship for support. LW acknowledges the support from DFG funding No. WA4313/1-1. Q.S. acknowledges the hospitality of University of California at Berkeley. 
\end{acknowledgments}

%

\end{document}